# AI on the Water: Applying DRL to Autonomous Vessel Navigation

Md Shadab Alam*[1], Sanjeev Kumar Ramkumar Sudha[2] and Abhilash Somayajula[3]

## ABSTRACT

Human decision-making errors cause a majority of globally reported marine accidents. As a result, automation in the marine industry has been gaining more attention in recent years. Obstacle avoidance becomes very challenging for an autonomous surface vehicle in an unknown environment. We explore the feasibility of using Deep Q-Learning (DQN), a deep reinforcement learning approach, for controlling an underactuated autonomous surface vehicle to follow a known path while avoiding collisions with static and dynamic obstacles. The ship's motion is described using a three-degree-of-freedom (3-DOF) dynamic model. The KRISO container ship (KCS) is chosen for this study because it is a benchmark hull used in several studies, and its hydrodynamic coefficients are readily available for numerical modelling. This study shows that Deep Reinforcement Learning (DRL) can achieve path following and collision avoidance successfully and can be a potential candidate that may be investigated further to achieve human-level or even better decision-making for autonomous marine vehicles.

## 1. INTRODUCTION

Recent International Maritime Organization (IMO) regulations on improving ship energy efficiency (measured by the Energy Efficiency Design Index- EEDI) and lowering CO emissions have pushed ship designers and owners to consider autonomous ships. Autonomous ships can increase fuel efficiency through better weather routing, improve safety with collision avoidance strategies, lead to efficient vessel operations, and be used for surveillance and reconnaissance missions. For example, after the deployment of the Yara Birkeland, the world's first autonomous and fully electric container vessel, in 2021, it was estimated that 40,000 trips performed by diesel vehicles annually could be eliminated. Similarly, up to 80-85 % of all reported marine accidents could be attributed to human error (Baker et al. 2005). These incidents endanger human lives and have the potential to cause serious harm to the environment. With the recent progress in artificial intelligence (AI), it is imperative to explore automation solutions for the maritime industry to reduce the probability of such incidents. AI advancements, particularly in reinforcement learning (RL), now provide new ways to explore ship trajectory tracking and collision avoidance solutions without human intervention.

[1] IIT Madras;  0000-0001-9184-9963;   1, oe21s007@smail.iitm.ac.in
[2] NTNU; 0000-0001-7442-3696;        2, sanjeev.k.r.sudha@ntnu.no
[3] IIT Madras; 0000-0002-5654-4627;   3, abhilash@iitm.ac.in

To track waypoints for path following, conventional autopilots use line of sight (LOS) guidance systems in conjunction with proportional-integral-derivative (PID) control systems (Lekkas et al. 2012, Moreira et al. 2007). However, today there is an increasing need to investigate AI-based control strategies, which can potentially perform better than traditional controllers. It has already been shown to be promising for specific applications such as active heave compensation (Zinage and Somayajula 2021, Zinage and Somayajula 2020) and dynamic positioning system (Lee et al. 2020). Some contemporary studies have also begun to investigate RL methods for path following and trajectory tracking. (Sivaraj et. al 2022) used Deep Q Network (DQN) for heading control and path following of a KVLCC2 ship under the influence of waves. (Woo et. al 2019) used a Deep Deterministic Policy Gradient (DDPG) algorithm-based controller for path following and successfully implemented on a full-scale Unmanned Surface Vessel (USV). (Meyer et. al 2020) applied the Proximal Policy Optimization (PPO) algorithm to demonstrate that an RL agent could successfully follow a predefined path while avoiding multiple obstacles.

## 2. THEORY

### 2.1. Ship Dynamic Model

The goal of waypoint tracking is to guide a vessel through a predetermined sequence of waypoints. Control laws are used to regulate the rudder angle according to a chosen guidance strategy, allowing the vessel to maintain a fixed propeller rotation rate. The problem is formulated using two coordinate systems: a global coordinate system (GCS) fixed to the Earth, where waypoint locations are specified, and a body coordinate system (BCS) fixed to the vessel.

The ship dynamics are mathematically modelled using the MMG (Maneuvering Modelling Group) model proposed in (Yoshimura and Masumoto 2012, Yasukawa and Yoshimura 2015). 3 non-linear equations of motion are used to solve for the vessel's motion in surge, sway and yaw directions. Based on a given rudder command $\delta_c$, the equations of motion are solved progressively at each time step as an initial value problem using a Runge-Kutta explicit solver. The kinematics model is represented by eq. 1.

$$\eta = [R(\psi)]\nu \qquad (1)$$

where, $R(\psi)$ represents the rotation matrix which transforms a vector from BCS to GCS.
Details of the modelling of propeller, rudder and hull hydrodynamic forces can be found in the works of (Yoshimura and Masumoto 2012, Deraj et al. 2023). All equations of motion are non-dimensionalized with prime-II system as described in (Fossen 2011), using the length between perpendiculars (L) and the design speed of the vessel (U).

## 2.2 Calculation of collision risk (CR)

When dealing with a scenario where a ship encounters multiple obstacles, it is crucial to evaluate the Collision Risk (CR) in order to determine an appropriate avoidance point for the autonomous navigation system. One method of evaluating CR is the Closest Point Approach (CPA), which utilizes the distance between the ship and obstacle at the closest point they approach each other while maintaining their current speed and direction. The Closest Point is defined as the Distance to the Closest Point Approach (DCPA), and the Time to the Closest Point Approach (TCPA) represents the duration for the obstacle to reach the Closest Point. In simpler terms, the DCPA gauges the seriousness of a potential collision while the TCPA represents the urgency of the situation, as depicted in Fig. 1.

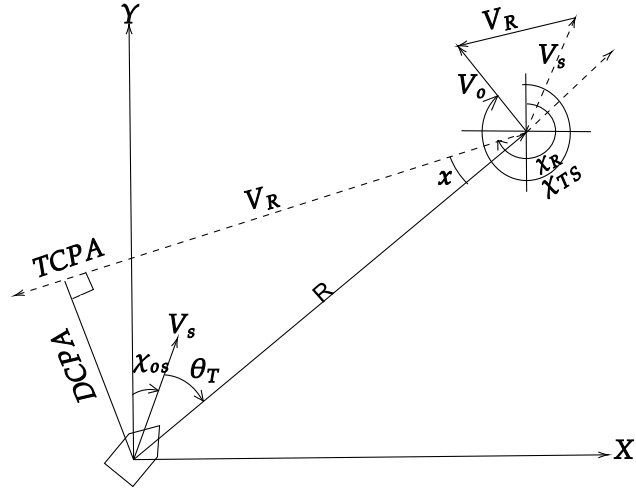

Figure 1. Collision Risk (CR)

Two different studies, (Mou et al. 2010) and (Zhen et al. 2017), presented techniques for evaluating CR by utilizing a combination of DCPA and TCPA. In the current research, CPA was employed to quantitatively evaluate both the ship and obstacle. Fig. 1 provides an explanation of the concepts of CPA, TCPA, and DCPA, and the equations presented in (4) is used to calculate TCPA and DCPA.

$$DCPA = R\sin(\chi_R - \chi_{os} - \theta_T - \pi) \qquad (4)$$
$$TCPA = \frac{R}{V_R}\cos(\chi_R - \chi_{os} - \theta_T - \pi)$$

where $R$ is the absolute distance between the ship and an obstacle, and $V_R$ and $\chi_R$ are the relative speed and course angle between them. In addition, $\chi_{OS}$ is the course of the obstacle, while $\theta$ is the bearing of the obstacle relative to the ship.

A number of research investigations have suggested techniques for quantitatively evaluating CR through the utilization of TCPA and DCPA. Among these, a straightforward evaluation equation was put forth by (Mou et. al 2010), which employs the exponential functions of TCPA and DCPA.

$$CR = \begin{cases} exp(-DCPA - TCPA), & if \ TCPA > 0 \\ 0 & , otherwise \end{cases} \quad (5)$$

The research examines the quantitative evaluation of the collision risk (CR) between the ship and the obstacle. The degree of danger in an encounter situation is determined by the values of DCPA and TCPA, which are included in Equation 5. A CR of 0 is recorded if DCPA or TCPA is infinite, indicating that there are no target obstacles near the ship or that TCPA is negative. If TCPA is negative, it means that either the ship and the obstacle have already passed each other or that they are moving in a way that avoids a collision.

## 2.3 Reinforcement Learning

The objective of an RL agent is to maximize a numerical signal, referred to as reward, which is provided by the environment based on the agent's actions (Sutton et. al 2018). In Deep Reinforcement Learning (DRL), the agent is a neural network that utilizes the environment's observations as input and produces actions as output. The agent's parameters are adjusted using an algorithm that relies on the rewards received from the environment to update the network's parameters. The role of rewards in DRL is crucial since the agent learns by trial and error. The agent explores various actions and observes the rewards received from the environment. Over time, the agent identifies which actions result in the highest rewards and updates its policy to maximize future rewards.

$$R = \sum_{t=0}^{T} r_t \quad (6)$$

## 3. IMPLEMENTATION OF DQN ALGORITHM FOR SHIP NAVIGATION

This section discusses the details of DQN algorithm (Mnih et al. 2013) applied to the autonomous ship for waypoint tracking and obstacle avoidance. This section describes how the observation state space, action space and rewards are defined to tackle this problem. Note that this study exclusively focuses on path following combined with obstacle avoidance. In this study, TensorFlow framework is used to model the RL agent. Fig. 2. shows the schematic representation of the problem statement.

## 3.1 Observation State Space

At each time step, the observation state vector represents the current state of the agent that is provided as an input to the Q-network. The RL agent decides what action to take based on this information. The observation state is defined using seven variables for static obstacle environment and nine for dynamic obstacle. The states for each environment are shown in Eq.8

$$s_{static} = [d_c, \chi_e, d_{wp}, r, d_{obs}, \psi_{obs}, S_{obs}]$$
$$s_{dynamic} = [d_c, \chi_e, d_{wp}, r, d_{obs}, \psi_{obs}, S_{obs}, v_x, v_y] \quad (8)$$

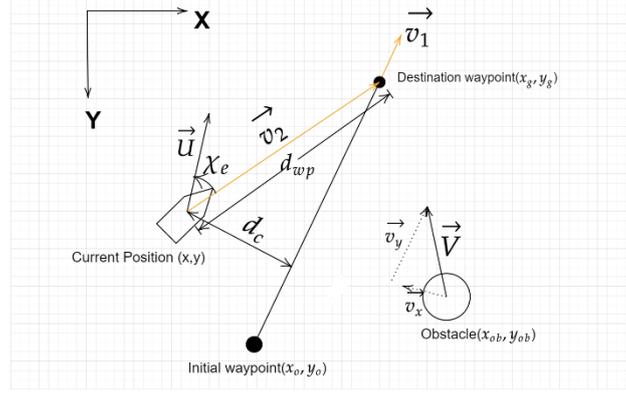

Figure 2. Representation of the problem statement

where $d_c$ is the cross track error, $\chi_e$ is the course angle error, $d_{wp}$ is the distance to waypoint, $r$ is the yaw rate, $d_{obs}$ is the distance to obstacle, $\psi_{obs}$ is the angle of the obstacle from ship, $S_{obs}$ is the size of the ship, $v_x$ is the relative velocity of the obstacle towards the ship with respect to ship and $v_y$ is the velocity of the obstacle perpendicular to the ship.

### 3.2    Action Space

The commanded rudder angle, $\delta_c$ is the action that the agent can choose at a given time step and it is divided into a set of five discrete values $\delta_c \in [-35°, -20°, 0°, 20°, 35°]$.

### 3.3    Reward Structure

The rewards must be designed in a way to help the agent achieve the required balance between the path following and obstacle avoidance objectives. The rewards obtained at any intermediate time step are given by (Deraj et al. 2023):

$$r_1 = 2 \exp\left(\frac{-d_c^2}{12.5}\right) - 1$$
$$r_2 = 1.3 \exp(-10|\chi_e|) - 0.3 \tag{9}$$
$$r_3 = \frac{-d_{wp}}{4}$$

where $r_1$ is the reward associated with cross track error, $r_2$ is the reward associated with course angle error. Finally to make overall reward negative, we introduced a reward $r_3$ that depends on the distance between the ship's current position and the destination waypoint. The reward at time step $r$ is denoted by $r_t$ and is given as the sum of the rewards from each component at that time step.

$$r_t = r_1 + r_2 + r_3$$
$$R = \sum_{t=0}^{n} r_t \tag{10}$$

In eq. 10, $r_t$ is the reward at time step $t$ and $R$ is the episode return, which is the cumulative sum of rewards obtained at each time step.

## 3.4 Training Process

During each training episode, the vessel commences at the origin point with a fixed initial velocity in the surge direction and an orientation along the positive X-axis of the Global Coordinate System ($\psi = 0$). The vessel's initial acceleration is zero in three degrees of freedom, while the initial velocity is zero in the sway and yaw motions. To enable sufficient exploration, the $\epsilon$-greedy strategy is used during training. The value of $\epsilon$ decreases linearly from 1 to 0 over the course of the training process, leading to increasing exploitation. The destination is randomly selected for each training episode, while the initial waypoint, velocity, and heading remain fixed at *(0,0)*, *1* and $0°$, respectively.

The distance between the destination and initial waypoints is randomly selected from a uniform distribution between *8L* and *18L*. For the static obstacle environment, in 60% of the training episodes, the obstacle is placed randomly within 0.25 to 0.75 times the distance between the waypoints, along the line joining the two points. Conversely, during the remaining 40% of the training iterations, the obstacle is placed randomly inside a circle with a radius ranging from 0.25 to 0.75 times the distance between the initial and goal waypoints. It is noteworthy that in this case, the obstacle may not be on the line joining the two points and may not even be in the ship's path. This biased placement of obstacles is intended to allow the agent to encounter obstacles in various scenarios and learn how to react when the obstacle is not in the vessel's path.

In the context of dynamic obstacle environment, a random destination waypoint is generated between *8L* to *18L*, and obstacles are randomly placed within *5L* to *20L*. Each obstacle has a velocity of *0* to *1.67U* and a size of *0* to *1L*. Incorporating the states of each obstacle would result in a large neural network size, so a collision risk calculation is proposed to identify the most critical obstacle that poses the highest threat of colliding with the ship. At each timestep, only information about the critical obstacle is passed to the neural network. By employing this approach, our study aims to improve the computational efficiency of the neural network in a dynamic obstacle environment.

In this task, each episode is limited to *160* time-steps, but it can terminate earlier if a termination condition is met. The success criterion for an episode is for the ship to enter a region within *0.5L* of the destination waypoint. If this condition is satisfied, a terminal reward of *+20* is awarded. On the other hand, if the ship collides with an obstacle, the episode ends, and the reward is *-100* in a static obstacle environment and *-200* in a dynamic obstacle environment. The discrepancy in the reward values is attributed to the smoother trajectory in a static environment, where only one obstacle can lead to a collision. However, in a dynamic environment, the obstacle's position changes due to its velocity, causing the critical obstacle for a collision to change as well. Therefore, the agent may take a longer route to avoid all obstacles and reach the destination safely. Another condition is established to determine whether the agent can still reach the destination point, or if the episode is a failure, and the ship is wandering aimlessly. The termination condition is as follows:

$$\overrightarrow{v_1} \cdot \overrightarrow{v_2} < 0 \text{ and } \overrightarrow{U} \cdot \overrightarrow{v_2} < 0 \tag{11}$$

## 4    HYPERPARAMETERS OF THE NETWORK

The training losses and episode returns averaged over 100 episodes are shown in Fig. 3 and Fig. 4. The learning rate is chosen as an exponentially decaying function. The hyperparameters for the model are given below in Table. 1.

TABLE I : Hyperparameters

| Hyperparameters | Value (Static) | Value (Dynamic) |
|---|---|---|
| Initial Learning rate | 0.00075 | 0.00075 |
| Decay Steps | 50000 | 50000 |
| Decay Rate | 0.4 | 0.5 |
| Hidden layers | 128,128 | 128,128 |
| Discount factor | 0.97 | 0.97 |
| Sample batch size | 128 | 128 |
| Replay buffer size | 100000 | 100000 |
| Activation function | Tanh | tanh |
| Number of episodes | 9000 | 8000 |
| Update frequency (time steps) | 10 | 5 |
| Target network update frequency (time steps) | 1 | 1 |
| Target update rate | 0.01 | 0.01 |

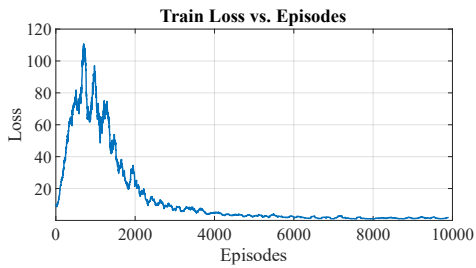
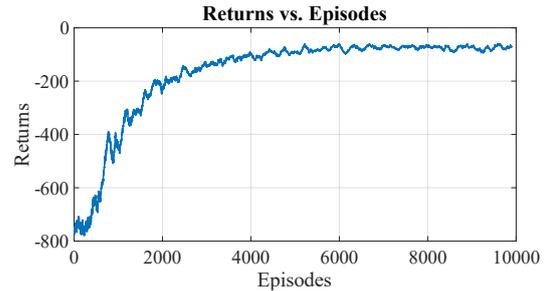

Figure 3. Training loss and returns for static obstacle environment

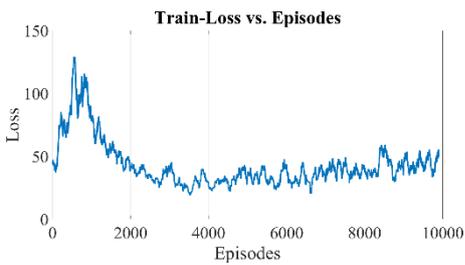
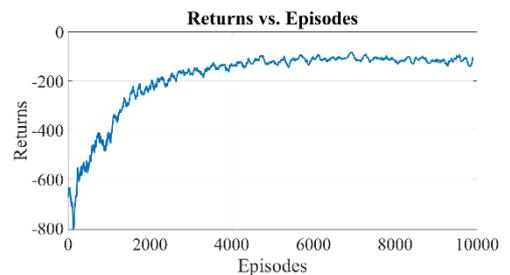

Figure 4. Training loss and returns for dynamic obstacle environment

## 5. RESULTS

A DQN agent is trained using the framework defined in Section. 3. The RL agent is tested by analyzing waypoint tracking and path following followed by adding obstacles between the defined path in different situations in a calm water scenario. A starting state is chosen where the vessel is oriented along the global X-axis and initialized with unit non-dimensional velocity in the surge direction. The ability of the agent to track points in different cases with the same initial starting state as mentioned above in each case.

Fig. 5(a) shows that the DQN based controller can maintain a heading, avoid an obstacle present on the desired path and regain heading. Fig. 5(b) and Fig. 5(c) demonstrate that the DQN agent can also track waypoints at a different heading than the initial heading and also avoid the obstacle present in between the path when the obstacle is present in between the waypoint. Fig. 6(a) and Fig. 6(b) demonstrate that the RL agent can successfully avoid the obstacle even in the case where the obstacle is not on the line joining the waypoints but within its natural path as it tracks the goal. It can observed that when the obstacle is large, the agent takes action to cross the line joining the waypoints and continue tracking goal when the distance to the obstacle is significantly safe.

The results depicted in [Link 1](Link 1), [Link 2](Link 2), [Link 3](Link 3) and [Link 4](Link 4) demonstrate that the DQN controller effectively navigates past dynamic obstacles and successfully reaches the intended destination.

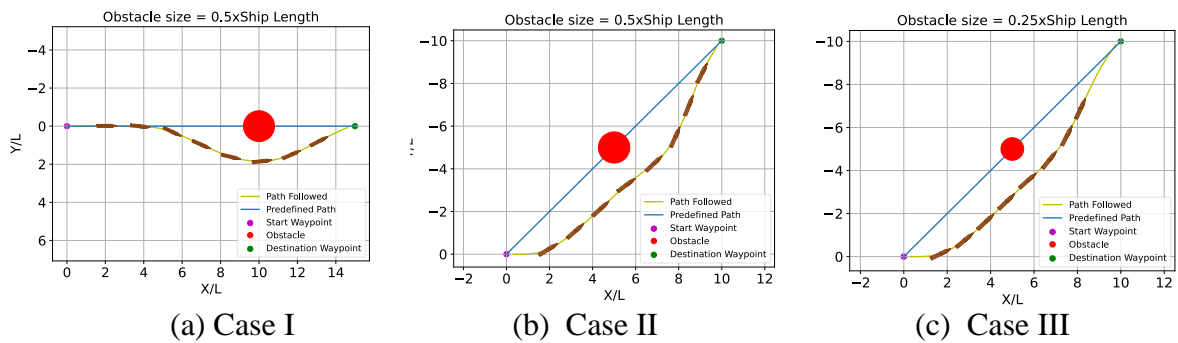

(a) Case I          (b) Case II          (c) Case III

Figure 5. Static Obstacle on the line joining the waypoints.

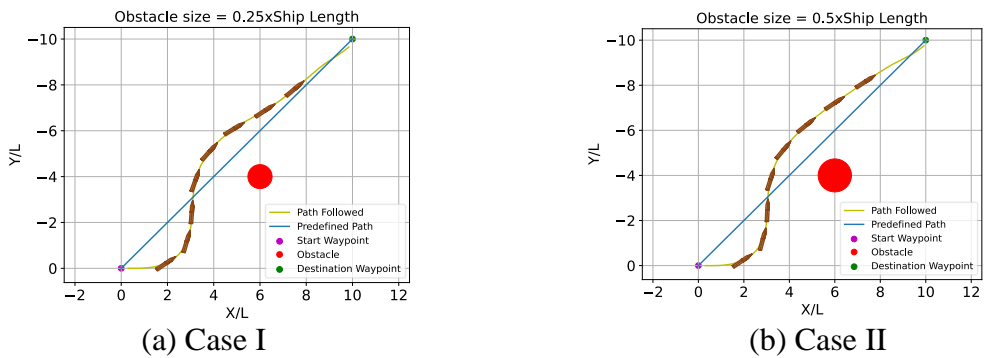

(a) Case I          (b) Case II

Figure 6. Static Obstacle on path of ships

## 6. DISCUSSIONS

The study evaluated the performance of a model trained on the static obstacle environment in scenarios with multiple obstacles. To test the model's ability to navigate through such environments, the authors employed the use of collision risk (CR), which was described in

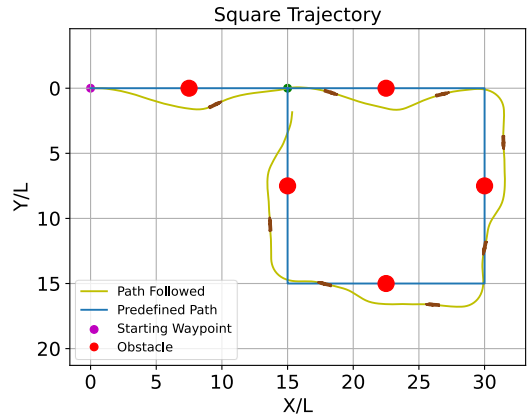

Figure 7. Square trajectory

Section 2.2. To calculate CR when dealing with static obstacles, the velocity of the obstacle was set to zero, while all other calculations remained unchanged. The results are presented in Fig. 7, which illustrates the agent's performance on a square trajectory with a side length of *15L* with an obstacle positioned between each waypoint. The findings revealed that the agent successfully avoided the obstacles and reached each waypoint along the trajectory.

## 7. CONCLUSION

This study has implemented a DRL based controller for an autonomous ship to navigate it successfully to the desired location while avoiding obstacles. The ability of the DQN agent to track waypoints in different scenarios and avoid static and dynamic obstacles is demonstrated.

Future work will include adding environmental effects due to currents and waves, which are important for meeting energy efficiency constraints set by the IMO. The RL agent performs well in simulations, but its practical effectiveness will be tested by implementing it on an ASV and comparing it to traditional controllers.

## 8. ACKNOWLEDGEMENT

This work was partially funded by the Science and Engineering Research Board (SERB) India - SERB Grant CRG/2020/003093 and New Faculty Seed Grant of IIT Madras. This work is also supported through the proposed {Center of Excellence for Marine Autonomous Systems (CMAS), IIT Madras} setup under the Institute of Eminence Scheme of Government of India.

**REFRENCES**